# Phase sensitive detection of extent of corrosion in steel reinforcing bars using eddy currents

Indrani Mukherjee, Jinit Patil, Sauvik Banerjee, and Siddharth Tallur

*Abstract—* **Corrosion of steel bars in reinforced cement concrete (RCC) structures leads to premature deterioration and increase in life cycle maintenance costs. Non-destructive testing (NDT) of incipient corrosion has been an impending task in this domain. We present a low cost sensing platform based on eddy current detection using anisotropic magnetoresistive (AMR) sensor to measure the extent of corrosion in steel reinforcing bars (rebars). The scheme employs phase-sensitive detection, wherein the phase shift in the sensor output (proportional to surface conductivity of the rebar) relative to the reference stimulus is measured by a lock-in amplifier and used to distinguish corroded and non-corroded rebars. The proof-of-concept sensor demonstration is able to resolve varying extents of rebar corrosion and can find potential applications as an NDT tool in a variety of industries.**

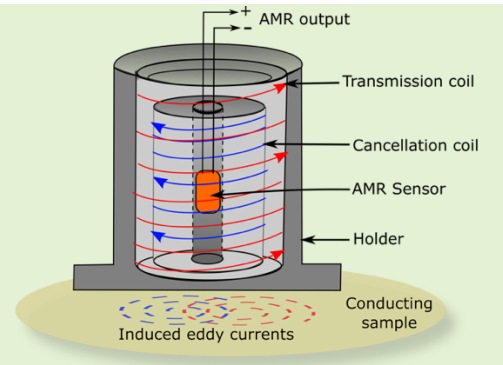

## I. INTRODUCTION

Corrosion of steel reinforcement is a global problem that leads to deterioration of RCC structures. Damages induced by steel corrosion result in repair and maintenance costs that exceed billions of dollars per year [1]. The costs vary depending on the condition of the concrete structures, including the cause of damage, degree of damage, and effect of damage on structural behavior. There are several techniques for Non-Destructive Testing (NDT) in Structural Health Monitoring (SHM) [2], [3], which are broadly captured in Table I. For effective SHM, detection of incipient corrosion in embedded reinforcement steel bars (rebars) is of utmost importance [4].

Eddy current based sensing is a good choice for conducting, ferromagnetic substances with high permeability [5]. Pulsed Eddy Current (PEC) based testing has an advantage over continuous wave eddy current owing to the ability to detect rebars buried several centimeters within concrete [6]–[8], albeit at the cost of expensive instrumentation. Moreover, PEC based testing can quantify volume loss in the rebar under test, and is capable of accurately quantifying corrosion in advanced stage of degradation of the structures [9]. Therefore, PEC is better suited to identify the location of corroded portions of rebars instead of preventive maintenance [10].

Conventional eddy current based testing uses alternating current (AC) signal to create a harmonic magnetic field using a solenoid. Due to mutual inductance of the solenoid and sample under test, the magnetic field produced by the eddy currents thus excited in the sample (typically metal) have a phase shift relative to the AC input to the solenoid. The phase shift is proportional to the conductivity of the sample [11], [12] and can thus be used to distinguish non-corroded rebars (higher conductivity) from corroded rebars (lower surface conductivity due to presence of rust i.e. iron oxide). Moreover, the eddy currents are localized to the external surface of the sample (limited by skin depth). Therefore, this technique can be used to quantify the extent of incipient corrosion or degradation of corrosion inhibiting coatings, by accurately recording the amplitude and phase shift in the eddy currents.

While the field generated by the eddy currents is detected with a receiver coil in conventional eddy current based testing, the sensitivity of the scheme is limited by the mutual and self inductance of the coils. Alternately, a highly sensitive magnetic field sensor may be incorporated in the coil assembly to detect minute changes in the field generated by eddy currents due to incipient corrosion. An anisotropic magneto-resistive (AMR) sensor having high sensitivity (typically order of few mV/V/Gauss used in a Wheatstone bridge configuration) can be used to detect weak magnetic fields [13], [14]. In this work, we demonstrate a phase sensitive detection scheme based on eddy current testing, utilizing an AMR sensor to detect changes in magnetic field strength due to corrosion in rebars. The amplitude and phase shift of the sensor output due to varying extents of corrosion are accurately measured with a lock-in amplifier, and well-resolved distinction between corroded and

Manuscript submitted for review on January 9th, 2020. This work was supported in part by the Government of India, Department of Science and Technology (DST) IMPRINT scheme under Grant IMP/2018/001442, and through a grant from Sanrachana Structural Strengthening Pvt. Ltd.

I. Mukherjee, J. Patil and S. Tallur* are with the Department of Electrical Engineering, IIT Bombay, Mumbai 400076 India (*email: stallur@ee.iitb.ac.in). S. Banerjee is with the Department of Civil Engineering, IIT Bombay, Mumbai 400076 India (e-mail: sauvik@civil.iitb.ac.in).

non-corroded areas of a rebar is observed in measurements.

TABLE I.  NDT TECHNIQUES FOR SHM

| NDT Techniques | Description | Advantages | Disadvantages |
|---|---|---|---|
| Optical | Visual inspection, fibre optics for NDT etc. | Non-contact, relatively inexpensive, large area coverage, fast | Low noise immunity, low penetration into insulating layer |
| Acoustic | Ultrasonic and acoustic emission based methods | Fast, large coverage, real time, deep penetration | Requires good acoustic coupling and smooth surfaces |
| Radio-graphic | X-rays or gamma rays for detection of faults and damages | Superior resolution, imaging possible | Expensive, requires access to both sides of the structure |
| Thermal | Thermo-graphic tests using infrared waves and eddy currents | Non-contact, large depth of penetration, imaging of micro-structure | Expensive and bulky, heat inducing setup, poor resolution on thick sections |
| Electro-chemical | Half-cell potential method, electro-chemical impedance spectroscopy | Fast, real-time measurement, provides degradation rate over time | Expertise needed to perform data analysis, unsuitable for in-situ detection |
| Electro-mechanical | Electro-mechanical impedance spectroscopy | Light weight, low power, detects cracks | Low sensitivity, limited to low actuation frequency |
| Electro-magnetic | Eddy current (pulsed and continuous), Magnetic Flux Leakage (MFL) | Relatively inexpensive, portable, high resolution, multilayer detection | Requires electrically conducting surfaces |

## II. METHODOLOGY

### A. Sensor Configuration

The sensing probe is designed to be portable, and in a form factor suitable for implementing an integrated scanning probe. Figure 1(a) shows all the components of the sensing assembly and their placements. The 3D printed holder houses two coils and the AMR sensor (Honeywell HMC1001). A solid cylindrical holder with a vertical slot designed to the dimensions of the AMR sensor is made into which the sensor can be inserted. Copper wire of diameter 0.05cm is wound on this cylinder (denoted as 'cancellation coil' in Figure 1(a)). This coil plus sensor setup is inserted into a larger hollow cylinder on which another coil is wound (denoted as 'transmission coil' in Figure 1(a)). This sub-assembly is then placed inside an external holder with a cylindrical cavity and square base ensuring that the sensitive axis of the AMR sensor is aligned with the axis of the two concentric coils by holding the entire sub-assembly in place. The number of turns of both the coils are designed so as to keep their turns-to-length ratio same. The net magnetic field at the AMR due to both coils can be zeroed by causing current to flow in clockwise direction in one coil and counter-clockwise direction in the other, and carefully adjusting their magnitudes to be equal. This results in reducing the AMR output offset voltage in standby condition i.e. in absence of a sample in the vicinity of the sensor. This method of compensating the offset voltage of the AMR sensor enables detection of small changes in the sensor output due to magnetic fields induced by the eddy currents in the conducting sample.

The parameters of the transmission coil are: resistance = $0.4\Omega$, inductance = $40\mu H$, outer diameter of cylinder = 2.5cm, length of the solenoid coil (L) = 2.5cm, diameter of wire = 0.05cm, and number of turns (N) = 45. The turns-to-length ratio is thus $N/L = 1800$. The cancellation coil has measured resistance = $0.1\Omega$, inductance = $7.5\mu H$, outer diameter of cylinder = 1.5 cm, L = 1.4 cm, and N = 25. The turns-to-length ratio is thus $N/L \approx 1800$. Since magnetic field inside a solenoid coil is directly proportional to its turns-to-length ratio, this ratio is kept the same by design for both coils to achieve magnetic field cancellation at the AMR sensor in absence of any external magnetic field.

### B. Experimental Setup

The test sample is prepared by artificially corroding a mild steel rebar to have three regions of visibly different corrosion levels. The specimen is shown in Figure 1(b). The region having the highest level of corrosion is marked as highly corroded followed by moderately corroded and uncorroded regions, with average diameters 20.61mm, 19.57mm and 17.85mm respectively, as measured with Vernier caliper. A wooden fixture is designed with a groove to hold the rebar firmly in place, and a slot along the length of the jig that allows smooth scanning of the assembly on the rebar with the square base of sensor holder fitting snugly in the slot (Figure 1(a)). This maintains the sensor alignment, such that the AMR sensor is directly atop the rebar while one slides the scanning probe along the length of the rebar. The clearance (lift-off) between the AMR sensor and the rebar is 2.5mm, unless specified otherwise. The sensor interface circuits are shown in Figure 1(c) and 1(d). Figure 1(c) consists of general purpose LM741 operational amplifiers used for driving current through the coils. The coils are represented by their equivalent circuit (series R-L circuit). The currents in the coils are adjusted by varying R1 and R3 (1kΩ potentiometers). Resistors R5 and R6 (220Ω each) are placed to limit the current in both coils. The maximum current that could be achieved with the circuit in Figure 1(c) is 10mA. This circuit shall be referred in subsequent sections as Circuit 1. Further, to increase the current in the coils in order to enhance the system's magnitude and phase response, a separate circuit is designed using power amplifiers LM380N. Figure 1(d) shows the circuit using general purpose LM741 operational amplifiers (op-amps) used to implement pre-amplifiers, and LM380N audio amplifiers to provide a large excitation current in both coils. The currents in the coils are adjusted by varying potentiometers R1 and R3 (5kΩ potentiometers). Current limiting power resistors R5 and R6 (100Ω, 2W each) limit the current in both coils to ~65mA rms. A dual channel DC power supply is used for powering the amplifier ICs and for providing bridge excitation voltage to the AMR sensor. For precision measurements, we use Stanford Research Systems SR530 lock-in amplifier in our experiments. It can generate reference AC signal through in-built function generator for exciting harmonic magnetic fields in both coils. The AC excitation frequency is 1kHz, and the AC signal excitation voltage provided by lock-in amplifier is 1V rms.

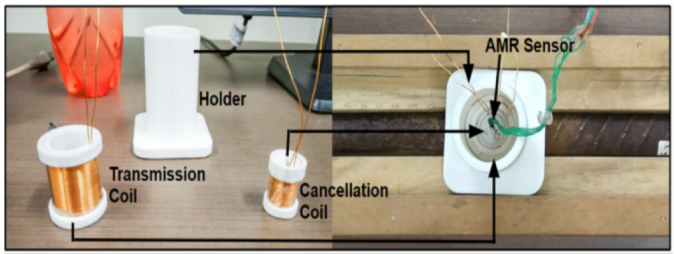
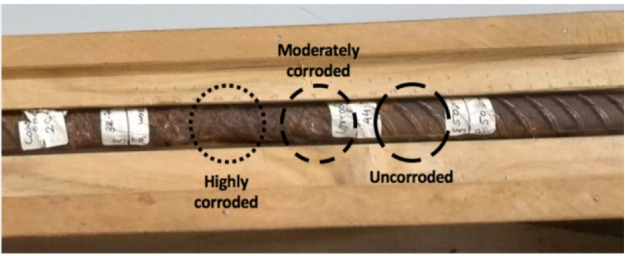

(a) (b)

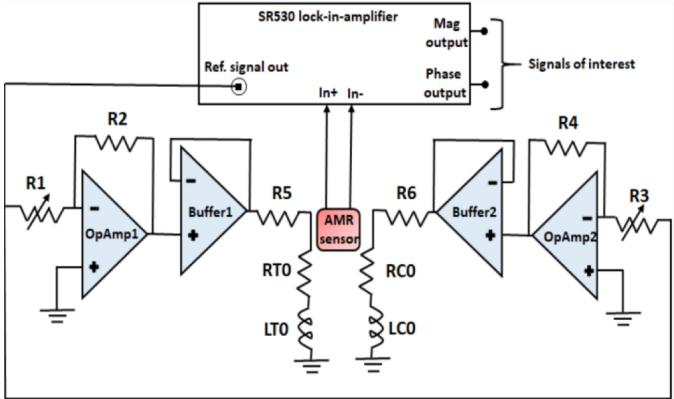
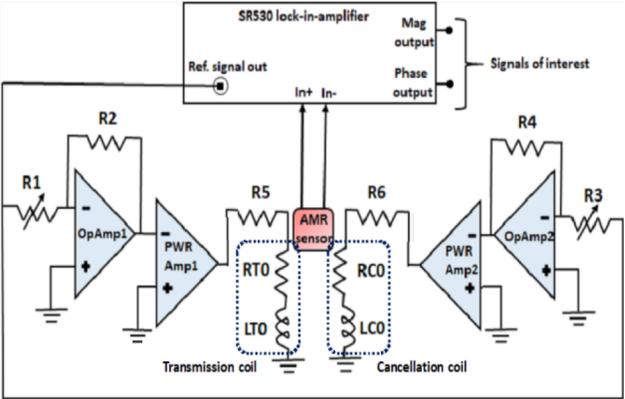

Circuit 1                                Circuit 2

(c)                                    (d)

Fig. 1. (a) Sensor assembly for corrosion monitoring in rebars. The left panel shows two co-axial coils and 3D printed holder that fit within each other as shown in the right panel. (b) The sample chosen for testing is a rebar of diameter 20mm. Three regions are chosen for testing as shown in the figure, with average diameter in uncorroded, moderately corroded and highly corroded sections being 20.61mm, 19.57mm and 17.85mm respectively. (c) Circuit diagram of the experiment using general purpose opamp LM741 for both pre-amplifier and buffer, the gains of which are controlled using potentiometers R1 and R3 giving maximum current in coils ~ 10mA. (d) Circuit diagram of the experiment using LM741 for pre-amplifier and LM380N audio amplifiers for generating a large excitation current (~65mA rms) in both coils. The SR530 lock-in amplifier provides the reference excitation voltage (marked as Ref. signal out) to both circuits (c) and (d), and measures the magnitude (amplitude) and phase shift of the differential output of the AMR bridge sensor with respect to the reference signal.

Figure 2 shows a photograph of the complete experimental setup. The AMR output is connected as differential input to the lock-in amplifier to obtain the voltage amplitude and phase shift corresponding to the field generated by the eddy currents. Using either circuit, the sensor system is scanned over the sample under test, and corresponding magnitude and phase outputs of the lock-in amplifier are recorded in all three regions. Ten measurements are conducted in each region.

## III. EXPERIMENTAL RESULTS AND DISCUSSION

The AMR offset voltage in absence of any metal in vicinity of the sensor is measured to be less than $10\mu V$, which is well within the dynamic range of the AMR sensor. The sensor is scanned over the rebar sample held inside the wooden fixture, and corresponding magnitude and phase outputs of the lock-in amplifier are recorded for all three regions of the sample under test shown in Figure 1(b). The above procedure is followed for both Circuit 1 and Circuit 2. Ten measurements are conducted in each of the three identified regions of different corrosion levels. Figure 3 shows the amplitude and normalized phase shift outputs of the lock-in amplifier, obtained for the three regions on the sample under test for three cases: (i) Circuit 1, lift-off = 2.5mm; (ii) Circuit 2, lift-off = 2.5mm; (iii) Circuit 2, lift-off = 1cm (with a 3D printed spacer inserted in the coil holder). The experiments are designed to observe the impact of coil current magnitude and lift-off on the sensor response. The measured phase of the output signal of the lock-in amplifier is normalized with respect to the mean of the phase output corresponding to measurements for the uncorroded region. The lock-in amplifier output for Circuit 1 is shown in Fig 3(a), while the lock-in amplifier outputs for Circuit 2 with 2.5mm lift-off and with 1cm lift-off are shown in Figures 3(b) and 3(c) respectively.

In Figure 3(a), the amplitude shows a decreasing trend whereas any inference cannot be drawn from the phase response. In Figure 3(b), while the amplitude shows a decreasing trend, the phase shift is observed to increase with extent of corrosion. Since corrosion leads to loss of ferromagnetic material and its conversion into iron oxide, these trends are intuitively justified. In Figure 3(c), while the amplitude can be resolved for the uncorroded and highly corroded regions, the same cannot be said for uncorroded and moderately corroded regions. The phase shift information is also inconclusive in this case. The improved response in Figure 3(b) is attributed to the larger coil currents and lower lift-off. As compared to Figure 3(a), increased coil current in Figure 3(b) leads to a larger magnitude of the field generated by the eddy currents. As compared to Figure 3(c), the lower lift-off in Figure 3(b) causes larger spatial overlap of the AMR sensor and the magnetic field lines generated by the eddy currents. Therefore, the scheme presented in this work is well-suited for assessing extent of corrosion in structures where small lift-off is permissible, such as sensors embedded within the infrastructure, or for applications such as pipelines and aircraft and ship bodies, or bridges.

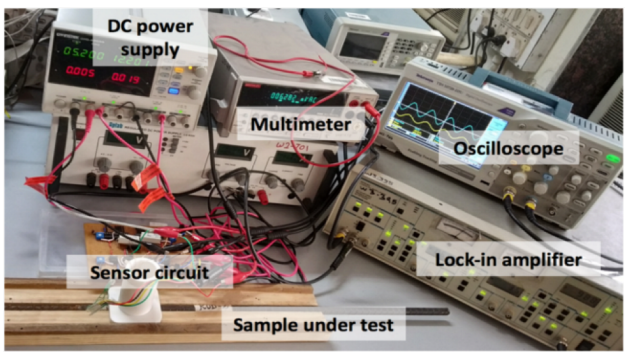

Fig. 2. Photograph of the experimental setup used for conducting measurements reported in this work.

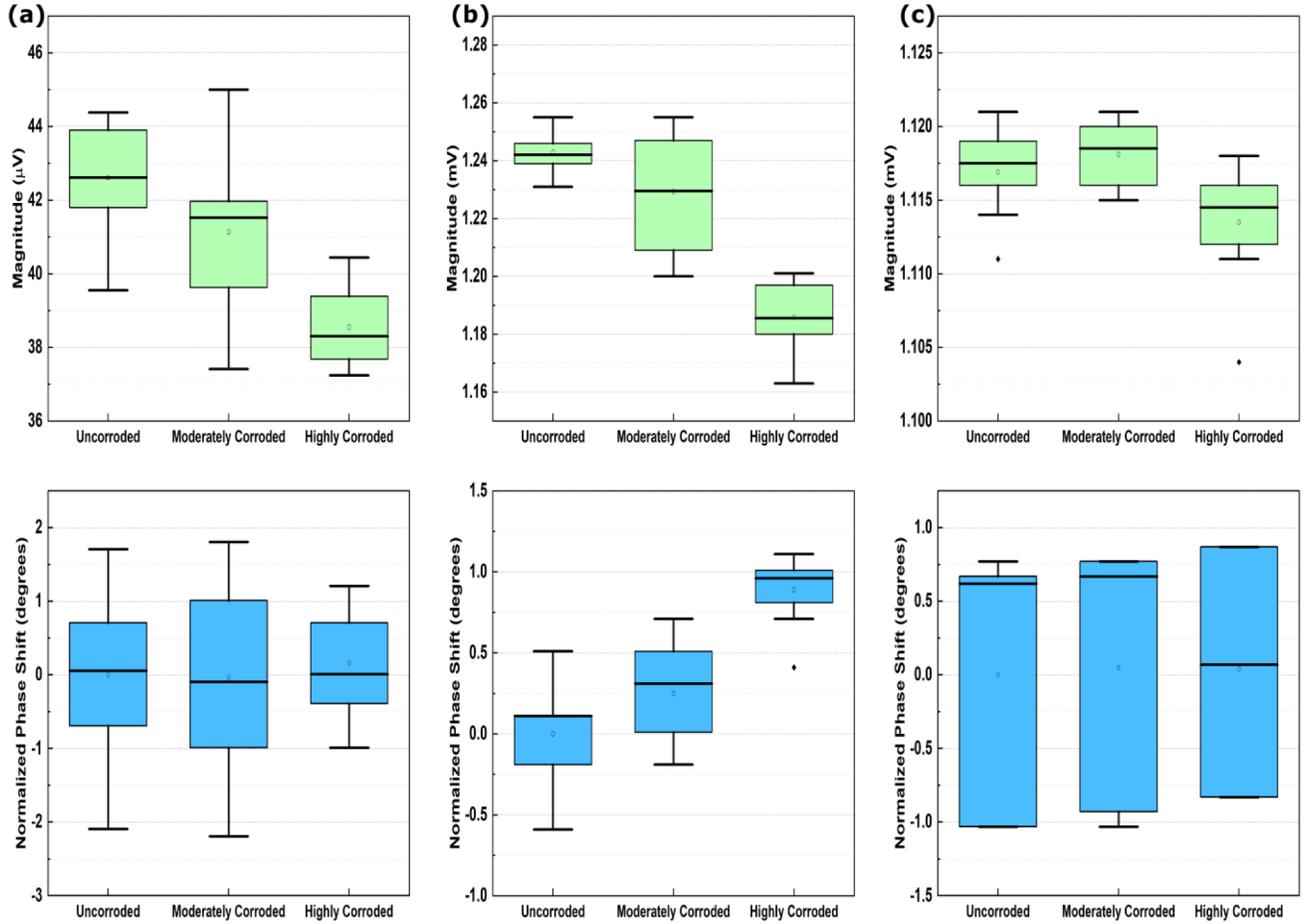

Fig. 3. Magnitude and normalized phase shift of the AMR sensor output when tested on the sample for three cases: (a) circuit 1 with 2.5mm lift-off, coil currents ≈ 10mA, (b) circuit 2 with 2.5mm lift-off, coil currents ≈ 65mA, (c) circuit 2 with 1cm lift-off, coil currents ≈ 65mA. Panel (b) shows well-resolved distinction between different levels of corrosion due to larger coil current and lowler lift-off.

## IV. Conclusion And Future Work

The principle of phase sensitive detection paired with the well-known technique of metal discrimination using eddy current has been utilized to ascertain the extent of corrosion in rebars. Corrosion being an extremely slow and gradual process, results in small changes in conductivity over long periods of time. Thus detection requires a sensor of high sensitivity and resolution leading to the choice of the AMR sensor as the prime sensing element. The custom sensor assembly is designed to enable a hand held, portable scanning probe that can distinguish between different levels of corrosion in a rebar. Phase sensitive detection using the precision lock-in amplifier provides simultaneous amplitude and phase information which can be used together for more accurate inference than amplitude data alone. This work provides a proof-of concept demonstration of the potential of the system to adequately resolve the different regions both in amplitude as well as phase of the sensor output with appropriate current levels for a given separation between the sensor and the sample. Due to cylindrical structure of the rod, a small fraction of the magnetic flux is cut by the AMR sensor, hence creating the need of a high power, high resolution phase sensitive detection system. As the distance between the sensor and the sample increases with lift-off, the sensing system's resolution degrades and higher coil current is needed to obtain satisfactory system performance. Nevertheless, the work presented here has applications for assessing the extent of degradation of exposed surfaces prone to corrosion, such as pipelines, ship and aircraft bodies etc. and also potentially for characterizing corrosion inhibiting coatings used for protecting such structures. Future work will be directed towards investigating feasibility of using this technique to identify incipient corrosion [15]–[17], optimizing the sensor assembly and the circuit to increase the coil current and sensor output, and miniaturizing the phase sensitive detector implemented by the lock-in amplifier to an on-board circuit [18]–[20] to enable a truly portable and field-deployable sensor system.

## ACKNOWLEDGMENT

The authors thank TATA Centre for assistance in 3D printing, and Prof. K. L. Narasimhan at IIT Bombay for providing guidance to the student authors on usage and handling of SR530. The measurements were carried out partly at Wadhwani Electronics Lab (WEL) and Applied Integrated Microsystems (AIMS) lab at IIT Bombay. Authors also thank Mr. Sitaram Varak at IIT Bombay for building the wooden test fixture.

## REFERENCES

[1] A. Sarja, "Reliability based life cycle design and maintenance planning," in *JCSS Workshop on Reliability Based Code Calibration*, 2002.

[2] P. J. Shull, *Nondestructive evaluation: Theory, techniques, and applications*, 1st ed. CRC Press, 2002.

[3] V. Giurgiutiu and A. Cuc, "Embedded non-destructive evaluation for structural health monitoring, damage detection, and failure prevention," *Shock Vib. Dig.*, vol. 37, no. 2, pp. 83–105, 2005, doi: 10.1177/0583102405052561.

[4] R. F. Muscat and A. R. Wilson, "Corrosion Onset Detection Sensor," *IEEE Sens. J.*, vol. 17, no. 24, pp. 8424–8430, 2017, doi: 10.1109/JSEN.2017.2764906.

[5] J. García-Martín, J. Gómez-Gil, and E. Vázquez-Sánchez, "Non-destructive techniques based on eddy current testing," *Sensors*, vol. 11, no. 3. pp. 2525–2565, 2011, doi: 10.3390/s110302525.

[6] N. P. de Alcantara, F. M. da Silva, M. T. Guimarães, and M. D. Pereira, "Corrosion assessment of steel bars used in reinforced concrete structures by means of eddy current testing," *Sensors (Switzerland)*, vol. 16, no. 1, 2015, doi: 10.3390/s16010015.

[7] Y. He, G. Tian, H. Zhang, M. Alamin, A. Simm, and P. Jackson, "Steel corrosion characterization using pulsed eddy current systems," *IEEE Sens. J.*, vol. 12, no. 6, pp. 2113–2120, 2012, doi: 10.1109/JSEN.2012.2184280.

[8] S. Giguère, "Pulsed eddy current: Finding corrosion independently of transducer lift-off," 2003, pp. 449–456, doi: 10.1063/1.1306083.

[9] S. Majidnia, R. Nilavalan, and J. Rudlin, "Investigation of an encircling pulsed eddy current probe for corrosion detection," in *Proceedings of IEEE Sensors*, 2014, vol. 2014-Decem, no. December, pp. 835–838, doi: 10.1109/ICSENS.2014.6985129.

[10] M. Fan, B. Cao, G. Tian, B. Ye, and W. Li, "Thickness measurement using liftoff point of intersection in pulsed eddy current responses for elimination of liftoff effect," *Sensors Actuators, A Phys.*, vol. 251, pp. 66–74, 2016, doi: 10.1016/j.sna.2016.10.003.

[11] S. Yamazaki, H. Nakane, and A. Tanaka, "Basic analysis of a metal detector," *IEEE Trans. Instrum. Meas.*, vol. 51, no. 4, pp. 810–814, 2002, doi: 10.1109/TIM.2002.803397.

[12] G. Miller, P. Gaydecki, S. Quek, B. T. Fernandes, and M. A. M. Zaid, "Detection and imaging of surface corrosion on steel reinforcing bars using a phase-sensitive inductive sensor intended for use with concrete," *NDT E Int.*, vol. 36, no. 1, pp. 19–26, 2003, doi: 10.1016/S0963-8695(02)00057-9.

[13] K. Tsukada *et al.*, "Detection of Inner Corrosion of Steel Construction Using Magnetic Resistance Sensor and Magnetic Spectroscopy Analysis," *IEEE Trans. Magn.*, vol. 52, no. 7, 2016, doi: 10.1109/TMAG.2016.2530851.

[14] D. F. He, M. Tachiki, and H. Itozaki, "Highly sensitive anisotropic magnetoresistance magnetometer for Eddy-current nondestructive evaluation," *Rev. Sci. Instrum.*, vol. 80, no. 3, 2009, doi: 10.1063/1.3098946.

[15] F. U. Renner, A. Stierle, H. Dosch, D. M. Kolb, T. L. Lee, and J. Zegenhagen, "Initial corrosion observed on the atomic scale," *Nature*, vol. 439, no. 7077, pp. 707–710, 2006, doi: 10.1038/nature04465.

[16] Z. Tian *et al.*, "Recent progress in the preparation of polyaniline nanostructures and their applications in anticorrosive coatings," *RSC Advances*, vol. 4, no. 54. pp. 28195–28208, 2014, doi: 10.1039/c4ra03146f.

[17] Y. Li, R. Hu, J. Wang, Y. Huang, and C. J. Lin, "Corrosion initiation of stainless steel in HCl solution studied using electrochemical noise and in-situ atomic force microscope," *Electrochim. Acta*, vol. 54, no. 27, pp. 7134–7140, 2009, doi: 10.1016/j.electacta.2009.07.042.

[18] J. Aguirre, N. Medrano, B. Calvo, and S. Celma, "Lock-in amplifier for portable sensing systems," *Electron. Lett.*, vol. 47, no. 21, pp. 1172–1173, 2011, doi: 10.1049/el.2011.2472.

[19] S. K. Sengupta, J. M. Farnham, and J. E. Whitten, "A simple low-cost lock-in amplifier for the laboratory," *J. Chem. Educ.*, vol. 82, no. 9, pp. 1399–1401, 2005, doi: 10.1021/ed082p1399.

[20] S. Carrato, G. Paolucci, R. Tommasini, and R. Rosei, "Versatile low-cost digital lock-in amplifier suitable for multichannel phase-sensitive detection," *Rev. Sci. Instrum.*, vol. 60, no. 7, pp. 2257–2259, 1989, doi: 10.1063/1.1140787.

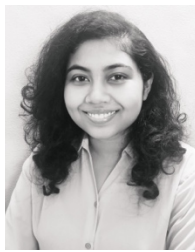

**Indrani Mukherjee** received the Bachelor of Technology degree in Electronics and Communication Engineering at Institute of Engineering and Management, Kolkata, India, in 2016. She is currently enrolled as a Master of Technology (M.Tech.) degree student in Electrical Engineering at IIT Bombay, specializing in Electronic Systems, and is working as a research assistant in AIMS Lab, IIT Bombay. Prior to joining IIT Bombay, she has worked as an Assistant System Engineer with Tata Consultancy Services, Kolkata, India (2016-18). Her research interest includes digital image processing, signal acquisition and compression and embedded systems.

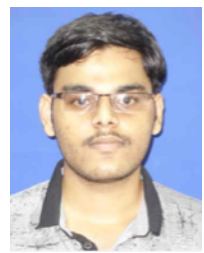

**Jinit Patil** received the Bachelor of Engineering degree in Instrumentation Engineering at Vidyavardhini's College of Engineering and Technology, Mumbai University, Mumbai, India, in 2017. He is currently enrolled as a Master of Technology (M.Tech.) degree student in Electrical Engineering at IIT Bombay, specializing in Electronic Systems. His research interests include embedded systems and robotics.

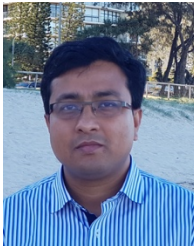

**Sauvik Banerjee** earned his PhD in Mechanical Engineering from University of California, Los Angeles (UCLA) in 2003, MTech in Structural Engineering from IIT Bombay in 2001, and BE in Civil Engineering from Bengal Engineering College, Shibpore in May 1999. He is currently a Professor of Structural Engineering in the Department of Civil Engineering, Indian Institute of Technology (IIT) Bombay. Prior to joining IIT Bombay, he had spent three years as an assistant professor at the Parks College of Engineering, Aviation and Technology, Saint Louis University, USA.

Dr. Banerjee's research focuses in the areas of structural health monitoring using wave propagation and vibration-based approaches, ultrasonic nondestructive evaluation of materials, and elastodynamic modelling of advanced composite structures. He has published over 100 articles in peer reviewed international journals and conference proceedings in these areas, including one book, two book chapters, and delivered invited talks in several important meetings. One of his PhD students has been awarded Excellence in Thesis Work by IIT Bombay in August 2017. He is recipient of Prof. S. P. Sukhatme Award for Excellence in Teaching, 2019 and Excellence in Teaching Award, 2013 at IIT Bombay.  In his early career, he has received Best Paper award at the SPIE conference on Smart NDE for Health Monitoring of Structural and Biological Systems, USA in 2003, the outstanding Doctor of Philosophy in Mechanical engineering for 2003-2004 at UCLA, the outstanding MTech in Civil Engineering for 2001 at IIT Bombay, and 'IIT-DAAD' German scholarship in 2000-2001. He is currently an associate editor of ASME Journal of Nondestructive Evaluation, Diagnostics and Prognostics of Engineering Systems.

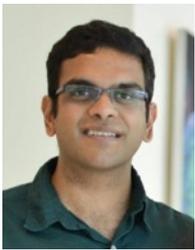

**Siddharth Tallur** received the B.Tech. degree in Electrical Engineering at IIT Bombay, Mumbai, India, in 2008, and M.S.-Ph.D. degree in Electrical and Computer Engineering at Cornell University, Ithaca NY, USA, in 2013. He is currently serving as an Assistant Professor in Electrical Engineering at IIT Bombay.

Dr. Tallur is an expert in physical and chemical sensor systems, MEMS and photonics, and embedded systems. His Ph.D. thesis research on low phase noise RF opto-mechanical and opto-acoustic oscillators won the best thesis award in the Electrical and Computer Engineering Department at Cornell University in September 2013. Following the completion of his Ph.D. in 2013, he worked at Analog Devices Inc. in Wilmington, MA, USA, where he conceived and led the characterization of novel gyroscope designs and mixed-signal circuit architectures for inertial motion-sensing applications. He joined IIT Bombay as Assistant Professor in Electrical Engineering in November 2016. He holds a visiting faculty appointment at the National Chiao Tung University (NCTU), Hsinchu, Taiwan and is currently serving as the faculty-in-charge at the Wadhwani Electronics Lab (WEL) at IIT Bombay.